\documentclass[letterpaper,twocolumn,showpacs,
prl,superscriptaddress,email]{revtex4}

\usepackage[pdftex]{graphicx}  
\usepackage{amssymb}
\usepackage{amsmath,amsfonts,latexsym}
\usepackage{array,tabularx,color}
\usepackage{dcolumn}               
\usepackage{bm}
\usepackage[latin1]{inputenc}
\usepackage{color}

\newcommand{\vect}[1]{{\mathbf #1}}

\newcommand{\Frac}[2]{\displaystyle\frac{#1}{#2}}

\begin{document}

\title{Ultra-Fast Stark-Induced Control of Polaritonic States}
\author{E.~Cancellieri}
\email[Corresponding author: ]{cancellieri@gmail.com}
\affiliation{Laboratoire Kastler Brossel, Universit\'{e} Pierre et Marie Curie,
Ecole Normale Sup\'{e}rieure et CNRS, Paris, France}
\author{A. Hayat}
\affiliation{Department of Physics, Centre for Quantum Information and Quantum Control,
and Institute for Optical Sciences, University of Toronto, Toronto, Ontario M5S 1A7, Canada}
\author{A. M. Steinberg}
\affiliation{Department of Physics, Centre for Quantum Information and Quantum Control,
and Institute for Optical Sciences, University of Toronto, Toronto, Ontario M5S 1A7, Canada}
\author{E. Giacobino}
\affiliation{Laboratoire Kastler Brossel, Universit\'{e} Pierre et Marie Curie,
Ecole Normale Sup\'{e}rieure et CNRS, Paris, France}
\author{A. Bramati}
\affiliation{Laboratoire Kastler Brossel, Universit\'{e} Pierre et Marie Curie,
Ecole Normale Sup\'{e}rieure et CNRS, Paris, France}
\date{\today}

\begin{abstract}
A laser pulse, several meV red-detuned from the excitonic line of a quantum well, has been shown to induce an almost instantaneous and rigid shift of the lower and upper polariton branches. Here we demonstrate that through this shift, ultra-fast all-optical control of the polariton population in a semiconductor microcavity should be achievable. In the proposed setup a Stark field is used to bring the lower polariton branch in or out of resonance with a quasi-resonant continuous-wave laser, thereby favoring or inhibiting the injection of polaritons into the cavity. Moreover we show that this technique allows for the implementation of optical switches with extremely high repetition rates.
\end{abstract}

\maketitle 

Systems made of quantum wells embedded in semiconductor microcavities, where light and matter are strongly coupled, provide a versatile area for the study of fundamental physics and new states of matter such as out-of-equilibrium polariton condensates. Due to the strong non-linear interaction between polaritons, they also have great potential for the implementation of next-generation all-optical computational technologies. Moreover, the very short lifetime of cavity-polaritons can in principle guarantee extremely fast operation rates. For these reasons, in recent years significant efforts have been devoted to the dynamical control of polaritonic systems and several proposals have been made to implement switches, spin-switches, transistors, and resonant tunneling diodes on cavity-polariton systems~\cite{degiorgi,gao_savvidis,paraiso,amo_switches,ballarini2013,adrados2010,adrados2011,nguyen2013,hayat2013}.

The general idea underlying all these proposals is to use exciton-exciton/polariton-polariton interactions to control the blueshift of the lower-polariton branch directly with the driving laser and, in this way, to manipulate the number of polaritons in a given spatial location and given energy state of a microcavity. The main disadvantage of this technique is that when the duration of the laser pulses used to trigger the desired operations is in the sub-picosecond regime, the corresponding bandwidth is of the order of the Rabi splitting of the system. Therefore, the effect of short pulses is not only to inject polaritons, but also to excite exciton reservoirs with relatively long lifetimes that relax into polaritons and considerably slow down the dynamics of the system. As a result, the dynamical control of polariton systems has not yet reached the ps or sub-ps regime.

Instead, the idea underlying the present work is to shift the energy of the polariton branches using the Stark effect produced by a laser far red-detuned from the excitonic line. It has been recently shown~\cite{hayat2012} that the dynamic Stark effect can be used to shift both the lower polariton (LP) and the upper polariton (UP) branches almost rigidly, with the shift lasting only for the duration of the Stark pulse (typically less than a ps). This ultra-fast control is possible because the Stark pulse does not excite reservoirs of long-lived particles.

In this letter, we propose a setup where a continuous-wave laser (CW) is used to resonantly inject polaritons into the microcavity and a Stark field is used to control the energy difference between the injecting laser and the LP branch. Once the parameters of the injecting laser (frequency, angle of incidence on the microcavity, and intensity) are fixed, the Stark intensity can be adjusted in order to take the lower polariton branch in or out of resonance with the frequency of the injecting laser, thereby increasing or decreasing the number of polaritons in the system, respectively. Here two cases are addressed: first the steady state case in which the Stark beam and the injection laser are CW and their intensities are changed adiabatically, and second the case in which a rapid change of the Stark field intensity, in the sup-picosecond range, is used to implement an ultra-fast all-optical polariton switch. When the Stark field takes the LP branch into resonance with the injecting laser, polaritons are efficiently injected in the cavity, and a bright ON state (corresponding to a high transmission from the microcavity) can be defined in contrast to a dark OFF state (corresponding to the case in which the LP branch and the injecting laser are out of resonance and the microcavity transmission is low).

We describe the system of a quantum well embedded in a semiconductor microcavity by means of a two-component wave function where the excitonic field of the quantum well ($\psi_X$) is strongly coupled to the photonic field confined in the microcavity ($\psi_C$) through the vacuum Rabi coupling $\Omega_R=d_{XC}\sqrt{2\omega_C/\hbar\epsilon_0V}$ between a dipole with matrix element $d_{XC}$ and the vacuum field in the cavity with volume V. The dynamics and steady state of the system are modeled by means of the generalized Gross-Pitaevskii equation~\cite{carusotto2005,cancellieri2010}:

\begin{equation}
i\hbar\partial_t \begin{pmatrix} \psi_X \\ \psi_C \end{pmatrix}
  =
\begin{pmatrix} 0 \\ F \end{pmatrix} + \left[\hat{H}_0
    + \begin{pmatrix} g_X|\psi_X|^2 & 0 \\ 0 &
      0 \end{pmatrix}\right]
\begin{pmatrix} \psi_X \\ \psi_C \end{pmatrix}
\nonumber
\end{equation}
\vspace{-0.5cm}
\begin{equation}
\hat{H}_0 =\hbar \begin{pmatrix}\omega_X^0-i\kappa_X & \Omega_R/2 \\
    \Omega_R/2 & \omega_C - i \kappa_C \end{pmatrix} .
\label{eq:model}
\end{equation}

Here a CW injection laser field (IN), nearly-resonant with the LP branch ($F=\hbar\sqrt{2\kappa_C}f({\bf r}) e^{i({\bf k}_{IN}\cdot{\bf r}-\omega_{IN} t)}$), injects polaritons with wavevector ${\bf k}_{IN}$ and frequency $\omega_{IN}$. Throughout the paper spatially homogeneous pumps $f({\bf r})=f$ are assumed, along with a flat exciton dispersion $\omega_X^0$ and a quadratic cavity dispersion $\omega_C({\bf k})=\omega^0_C- \frac{\hbar^2{\bf k}^2}{m_C}$, with $m_C=2.3\times10^{-5}m_0$ and $m_0$ is the electron mass. The vacuum Rabi frequency is set to $10.0$~meV, $\kappa_X$ and $\kappa_C$ are the excitonic and photonic decay rates. The exciton-exciton interaction strength $g_X$ is of the order of 40 $\mu$eV$\mu m^2$~\cite{amo2009}, and for the sake of generality we rescale the fields $\psi_{X,C}\rightarrow \psi_{X,C}\sqrt{g_X}$ and the pump field $f\rightarrow f\sqrt{g_X}$. For the implementation of the proposed setup, a value of $g_X$ of 40 $\mu$eV$\mu m^2$ implies powers of the quasi-resonant laser in the range of 100 $mW$ for a spot of $100\times 100$ $\mu^2$. Throughout the paper the zero energy is set to the bare exciton frequency and the exciton-photon detuning is taken to be equal to zero ($\omega_X^0=\omega_C({\bf k}=0)$). Due to the Rabi coupling $\Omega_R$, the two eigenmodes of $\hat{H}_0$ are the LP and UP branches:
\begin{equation}
\!E_{LP,UP}({\bf k})\!=\!\frac{\hbar}{2}\!\left[\omega_X^0\!+\!\omega_C({\bf k})\!\pm\!\sqrt{\Omega_R^2\!+\![\omega_X^0\!-\!\omega_C({\bf k})]^2}\right]\!.\!\!\!\!
\label{eq:lpupenergy}
\end{equation}

\begin{figure}
  \centering
  \includegraphics[width=1.00\linewidth]{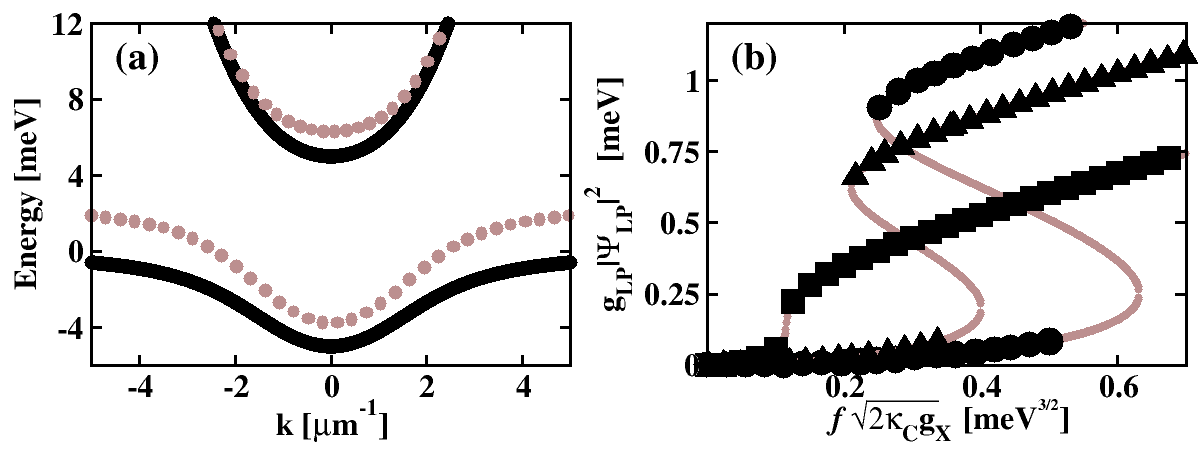}
  \caption{\label{fig:newLPUP}(Color online) (a) LP and UP branches as a function of the wavevector $k_x$ with $k_y=0$. The bare case (solid black) is compared with the dressed case obtained fully diagonalizing eq. (\ref{eq:modelstark}) with $\hbar\Omega_p=15$ meV (dotted brown). (b) Mean field polariton energy as a function of the quasi-resonant injecting laser intensity $f$ when $\vect{k}_{IN}=0$ $\mu m^{-1}$ and $\Delta=0.7$ meV. The stable solutions of the systems are represented for three Stark fields $\hbar\Omega_p=0.0,6.0$ and $10.0$ meV (circles, triangles, and squares) for linewidths $\hbar\kappa_C=\hbar\kappa_X=0.10$ meV. The brown curves indicate the unstable solutions of the system.}
\end{figure}

Throughout this Letter the injecting laser frequency will be assumed to be slightly blue detuned with respect to the LP mode: $\Delta=\omega_{IN}-E_{LP}(\vect{k}_{IN})>0$. The effect of a Stark field coupled to a quantum well exciton embedded in a microcavity can be fully described by a three dimensional Hamiltonian including the excitonic and photonic components and the non-resonant Stark pump:
\begin{equation}
H =\hbar
\begin{pmatrix} \omega_X^0 & \Omega_R/2 & \Omega_p/2 \\
                \Omega_R/2 & \omega_C({\bf k}) & 0 \\
                \Omega_p/2 &      0     & \omega_p 
\end{pmatrix}\; ,
\label{eq:modelstark}
\end{equation}
\noindent
where the decay rates have been omitted for the sake of clarity and where $\hbar\omega_p=-50$ meV (below the excitonic line), is the frequency of the Stark laser and $\Omega_p=d_{Xp}|\varepsilon_p|$ is the Rabi coupling between the exciton and the Stark field, with dipole matrix element $d_{Xp}$ and electric field $\varepsilon_p$. The full diagonalization of equation (\ref{eq:modelstark}) gives two new blueshifted LP and UP modes since the Stark frequency is red detuned with respect to the excitonic frequency. As shown in figure~\ref{fig:newLPUP}(a), for $\hbar\Omega_p=20$ meV the two dressed states are blueshifted by about $2.5$ meV with respect to the bare LP and UP, at ${\bf k}=0$. For weak Stark intensities it has been shown~\cite{mysyrowicz1986} that the dressed polariton modes are well approximated by the polariton modes obtained from equation~(\ref{eq:lpupenergy}) with the excitonic line blueshifted as:
\begin{equation}
\omega_X(\Omega_p,\omega_p)=\frac{1}{2}
\left(\omega_X^0+\omega_p+\sqrt{(\omega_X^0-\omega_p)^2+\Omega_p^2}\right).
\label{eq:excitonshift}
\end{equation}
\vspace{0.0cm}

This approximation, valid for blueshifts smaller than about 1 meV, considerably simplifies the analytical treatment of the system and will be adopted throughout the rest of this Letter. In order to understand the underlying mechanism governing the system, its steady state stable solutions are evaluated following the same perturbative Bogoliubov-like analysis already used for resonantly pumped polaritons in Refs.~\cite{carusotto2004,carusotto2005,cancellieri2010}. In the present case, however, the excitonic energy is a function of the parameters of the Stark field as described in equation~(\ref{eq:excitonshift}). Within this approach the homogeneous profile of the pump determines the spatially homogeneous stationary states $\psi_{X,C}^{(0)}e^{-i(\omega_{IN}t-\vect{k}_{IN}\cdot\vect{r})}$ of the exciton and photon, by means of the mean-field equations:
\begin{align}
\nonumber
0&=\!\!\left[\omega_X(\Omega_p,\omega_p)-\omega_{IN}-i\kappa_X+\frac{g_X}{\hbar}
    |\psi_{X}^{(0)}|^2\right]\!\psi_{X}^{(0)}+\Frac{\Omega_R}{2}
  \psi_{C}^{(0)}\\
f&\sqrt{2\kappa_C}\!=\!-\!\left[\omega_C({\bf k}_{IN})-\omega_{IN}-i\kappa_C\right]
  \psi_{C}^{(0)} - \Frac{\Omega_R}{2} \psi_{X}^{(0)}\; .
\label{eq:meanfield}
\end{align}

In order to evaluate the stability of the solutions of eq.~(\ref{eq:meanfield}), fluctuations around the mean-field state are introduced and both exciton and photon fields are expanded above their mean-field homogeneous stationary states as: $\psi_{X,C}(\vect{r},t)=e^{-i\omega_{IN}t}\left[e^{i\vect{k}_{IN}\cdot\vect{r}}\psi_{X,C}^{(0)}+\delta\psi_{X,C}(\vect{r},t)\right]$. As has been shown~\cite{carusotto2004,carusotto2005} the fluctuations above the stationary state can be re-written in terms of particle-like ($u_{X,C}$) and hole-like ($v_{X,C}$) excitations: $\delta\psi_{X,C}(\vect{r},t)\!=\sum_{\vect{k}}\left(e^{-i \omega t}e^{i \vect{k}\vect{r}}u_{X,C;\vect{k}}+e^{i\omega t}e^{-i(\vect{k}-2\vect{k}_{IN})\vect{r}}v_{X,C;\vect{k}}\right)$, and the stability of the system obtained studying the imaginary part of their spectrum~\cite{carusotto2004}. The spectrum needed for this evaluation is obtained simply by solving the eigenvalue equation: $\left[(\omega+\omega_{IN})\mathbb{I}-\mathbb{L}\right]\begin{pmatrix}u_{X;\vect{k}} & u_{C;\vect{k}} & v_{X;\vect{k}} & v_{C;\vect{k}}\end{pmatrix}^T =0$, where $\mathbb{I}$ is the identity matrix and $\mathbb{L}$ is the matrix

\begin{widetext}
\begin{equation*}
\begin{pmatrix}
  \omega_X(\Omega_p,\omega_p) + 2\frac{g_X}{\hbar}|\psi_X^{(0)}|^2 -i\kappa_X &
  \Omega_R/2 & \frac{g_X}{\hbar} {\psi_X^{(0)}}^2 & 0\\
  \Omega_2/2 & \omega_C({\bf k}) -i\kappa_C & 0 & 0\\
  -\frac{g_X}{\hbar} {{\psi_X^{(0)}}^2}^* & 0 & 2\omega_{IN} - \omega_X(\Omega_p,\omega_p) - 2\frac{g_X}{\hbar}
  |\psi_X^{(0)}|^2 -i\kappa_X & -\Omega_R/2 \\
  0 & 0 & -\Omega_R/2 & 2\omega_{IN} - \omega_C(2{\bf k}_{IN} - \vect{k})
  -i\kappa_C
\end{pmatrix} \! .
\end{equation*}
\end{widetext}

The stability curve of the system, in the case of two CW laser fields, is plotted in Figure~\ref{fig:newLPUP}(b) as a function of the strength of the quasi-resonant pump and for three different values of the Stark field. Here and in the following, since we are considering ${\bf k}_{IN}=0$, we plot the mean field polariton energy $g_{LP}|\psi_{LP}|^2$ using the approximations: $g_{LP}=g_X/4$ and $\psi_{LP}=\frac{1}{\sqrt{2}}(\psi_X-\psi_C)$. Two features can be noticed in this plot: first, the width of the bistable region decreases until it disappears above a critical Stark intensity. This decrease is related to the fact that the width of the bistable region is proportional to the detuning between the injecting laser and the LP branch at $f=0$, and this detuning decreases for higher Stark fields. Secondly, when the system is highly populated (upper part of the stability curve), the number of polaritons in the cavity for fixed intensity $f$ is lower for stronger Stark fields. This can be understood considering that the number of polaritons in the cavity when the LP branch is in resonance with the injecting laser is proportional to the detuning between the injecting laser and the bare LP branch. The effect of the Stark field is to decrease this detuning by independently dressing the LP branch, and therefore the number of polaritons in the ON state must be smaller.

\begin{figure}
  \centering
  \includegraphics[width=1.0\linewidth]{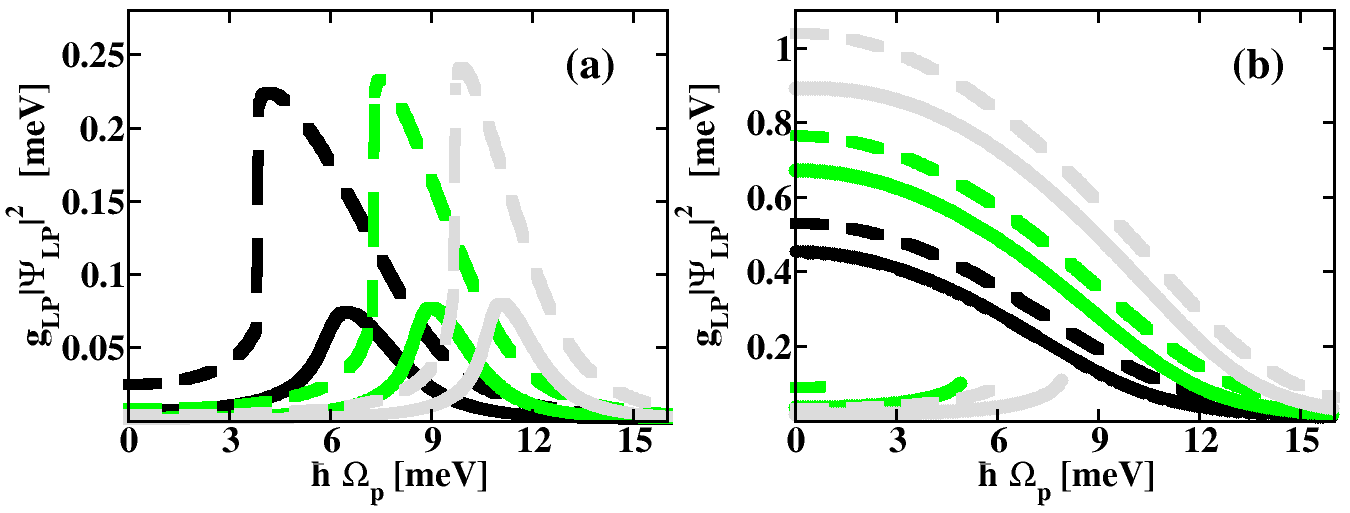}
  \caption{\label{fig:pop_vs_stark}(Color online) Mean field polariton energy as a function of the Stark field intensity $\Omega_p$. The injecting laser is taken to have $\vect{k}_{IN}=0$ $\mu m^{-1}$ and the three cases (black, green and grey) correspond to $\Delta=0.3,0.5$ and $0.7$ meV (from left to right in (a) and from bottom to top in (b)). (a) first class of switches with injecting laser intensity $f\sqrt{2\kappa_Cg_X}=0.075, 0.130$ meV$^{3/2}$ (solid and dashed lines respectively). (b) second class of switches with injecting laser intensity $f\sqrt{2\kappa_Cg_X}=0.25, 0.30$ meV$^{3/2}$ (solid and dashed lines respectively). For $\Delta=0.5$ and $0.7$ states at low density are present in (b) since for these sets of parameters the system display bistability at $\Omega_p=0$. The linewidths are $\hbar\kappa_C=\hbar\kappa_X=0.10$ meV. Note that in the $\Delta=0.3$ case no low-density stable solutions are available at low $\Omega_p$ since, in this case, the intensities $f$ considered are above the bistability threshold.}
\end{figure}

Still in the steady-state case, in order to get a better insight on the effect of the Stark field, instead of fixing $\Omega_p$ and changing $f$, it is possible to study the stability of the system in the opposite way by fixing the injecting laser intensity $f$ and changing the Stark intensity $\Omega_p$. We study this in two cases: weak and strong intensities of the quasi-resonant injecting laser, plotted in figures~\ref{fig:pop_vs_stark}(a) and (b) respectively. As will be explained below, these two cases correspond to two kinds of switches that can be implemented using the Stark effect.

In the first case few polaritons are present in the cavity when $\Omega_p=0$ since the injecting laser is weak and $\Delta>0$, and therefore, it is possible to define an OFF state for the corresponding switch. When $\Omega_p$ is increased, the LP branch shifts towards higher energies until it becomes resonant with the injecting laser, at a given value $\Omega_p=\Omega_p^{ON}$. At this point, polaritons are efficiently injected into the cavity and the switch turns ON. Now, if $\Omega_p$ is kept fixed at $\Omega_p^{ON}$ the switch remains in the ON state, if, instead, $\Omega_p$ is decreased or further increased, the switch turns OFF since the LP and the injecting laser are no longer in resonance. This behavior is clearly visible in the peak-like shape of the polariton density plotted as a function of $\Omega_p$ in fig.~\ref{fig:pop_vs_stark}(a). As also shown in fig.~\ref{fig:pop_vs_stark}(a), the switching-ON intensity $\Omega_p^{ON}$ increases with higher detuning $\Delta$ of the injecting laser with respect to the bare LP branch (curves with different colors). This is because higher $\Omega_p$ are needed to blueshift the LP to higher values and to reach the frequency of the injecting laser. The shifts considered here, between 0.3 and 0.7 meV, have been experimentally achieved for values of the Stark pump fluence between 0.75 and 2.0 mJ/cm$^2$~\cite{hayat2012}. This confirms the experimental feasibility of this kind of switching operation. At this point it is worth noting two things. First, since the injecting laser intensity $f$ is weak, the peak density of the ON state is weakly dependent on the initial detuning as when the LP and the injecting laser are in resonance, the number of polaritons that are injected into the cavity depends only on $f$ and on the polariton lifetime. Second, the $\Omega_p^{ON}$ at which the polariton intensity peaks shifts slightly towards lower values when the pump intensity $f$ is increased (change from solid to dashed lines with same color). This can be understood observing that the LP branch is shifted by both the Stark field and by the polariton population in the cavity. Therefore, if the polariton density is higher, the LP is in resonance with the injecting laser at smaller Stark intensities.

In the second case, when the injecting laser is strong, the corresponding switch, contrary to the first case, lies in an ON state when $\Omega_p=0$ and is turned OFF at high Stark field intensities. As before, the frequency of the injecting laser is quasi-resonant and blue detuned ($\Delta>0$) from the bare LP branch but, in this case, the pump intensity $f$ is strong enough to blueshift the LP branch into resonance. As shown in fig.~\ref{fig:pop_vs_stark}(b), when $\Omega_p$ increases the switch is turned OFF since, due to the Stark field, the LP branch is taken out of resonance with the injecting laser. When $\Omega_p$ is decreased back to zero the polariton branch is shifted back into resonance with the injecting laser and the switch turns ON again. While in the previous case the effect of the Stark field was to bring the LP from values below the energy of the injecting laser into resonance with it, here the device works exactly in the opposite way: the LP branch starts in resonance with the injecting laser and the Stark field takes it out of resonance. As for the first class of devices, the switching intensity ($\Omega_p^{OFF}$ in this case rather than $\Omega_p^{ON}$) is higher for higher detunings $\Delta$, since the polariton branches have to be shifted more. However, unlike the previous case, here the peak intensity of the ON state depends strongly on the detuning $\Delta$ (curves with different colors) since this intensity determines the polariton density that the quasi-resonant laser has to inject into the cavity in order to keep the LP dressed and in resonance. It is also worth noting that in this case, once the intensity $f$ is sufficiently high to sustain the ON state at $\Omega_p=0$, the intensity of the ON state depends only weakly on a further increase of $f$ since the system is in an optical limiter regime (solid and dashed curves with same color).

\begin{figure}
  \centering
  \includegraphics[width=1.00\linewidth]{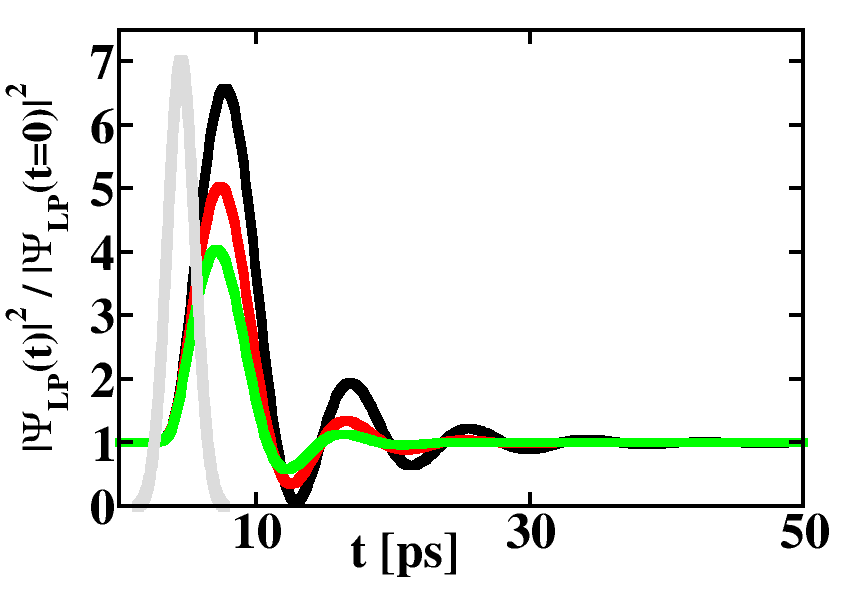}
  \caption{\label{fig:dynamics}(Color online) Time evolution of the polariton population in the cavity after the arrival of a Stark pulse (grey line) with values normalized to the population before the arrival of the pulse. The green, red and black cases correspond to plariton lifetimes of $2/(\kappa_C+\kappa_X)=3.3, 4.4$, and $6.6$ ps. The detuning between the CW laser and the LP branch is $\Delta=0.5$ meV, the CW pump intensity is $f\sqrt{2\kappa_Cg_X}=0.13$ meV$^{3/2}$, and the Stark intensity is $\hbar\Omega_p=20$ meV. The dashed line schematically represent the Stark pulse.}
\end{figure}

Finally, we have studied the dynamics of the system for the first class of switches described above, by numerically solving the radial component of the Gross-Pitaevskii equation (note that due to the circular symmetry of the infinite and homogenous system the radial component of the wavefunction carries all the information needed for the full solution of the problem). Figure~\ref{fig:dynamics} shows the response of a system, initially in its steady state, after the arrival of a Gaussian (in time) Stark pulse ($\sigma_t=1.0$ ps), for three different polariton lifetimes. The arrival of the Stark pulse (grey line in fig.~\ref{fig:dynamics}) triggers the blueshift of the LP branch towards the resonance with the CW laser, thereby enhancing the polariton population. After the end of the Stark pulse, the polariton population decreases back to its original steady state value displaying fast oscillations with a period of about $8$ ps that corresponds to the detuning ($\Delta=0.5$ meV) between the injecting laser and the bare LP branch. These oscillations are the same as those displayed by the system when a laser field is suddenly turned on (step-like) on an empty microcavity, and their duration is proportional to the polariton lifetime. First of all it should be noted that while the switching ON of the device is extremely fast, since it is determined by the rate at which photons are transformed into polaritons (the Rabi frequency $\Omega_R$), the switching OFF is determined by the polariton lifetime, since the system goes back to its initial condition as polaritons decay. Therefore longer polariton lifetimes (black curve), slowing down the process of restoring of the initial condition, decrease the repetition rate at which the device can work. On the other hand, longer lifetimes allow for more efficient polariton injection, and yield higher polariton densities in the ON state. This is a principal figure of merit for switches since higher brightness ratios (population ON/population OFF) lead to better device performance.

In conclusion, we have shown that a Stark field far red-detuned from the excitonic resonance of a semiconductor microcavity can control the number of polaritons injected into the cavity by a quasi-resonant CW laser. In this way two classes of switches can be defined depending on whether the ON state corresponds to low or high Stark field intensities. Interestingly, we have also demonstrated that the switch between the ON and OFF states can be performed with repetition rates never reached to date in these systems. In fact, the injection of polariton is limited solely by the vacuum Rabi frequency, while the resetting of the initial condition depends only on the time needed for polaritons to decay. This technique not only allows for the implementation of switches with high modulation depth, but may also be used to study the response of a polaritonic fluid to fast control fields.

The authors acknowledge the ANR Quandyde, NSERC, and CIFAR projects; F. M. Marchetti, M. H. Szymanska and F. P. Laussy for helping the code development; C. Tejedor for the use of the computational facilities of the UAM; D. Ballarini and D. Sanvitto for fruitful discussions.

\bibliography{mybiblio}{}

\end{document}